\documentclass[final,5p,twocolumn]{elsarticle}

\usepackage{amssymb}

\usepackage{graphicx}
\usepackage{xcolor}
\usepackage{layouts}
\usepackage{multicol}
\usepackage{cuted}
\usepackage{threeparttable}
\usepackage{booktabs}
\usepackage{longtable}
\usepackage{hyperref}
\usepackage{amsthm, amsmath}
\usepackage{algorithm}
\usepackage{algpseudocode}
\usepackage{placeins}
\usepackage[symbol]{footmisc}
\usepackage{amsfonts}
\usepackage{pgfplots}
\pgfplotsset{compat=1.18}
\usepackage{subcaption}

\newcommand{\beq}{\begin{equation}}
\newcommand{\eeq}{\end{equation}}

\def\ten   #1{\boldsymbol{#1}}
\def\vec   #1{\boldsymbol{#1}}

\graphicspath{{./}{../}{../figures/}}
\journal{arXiv}

\begin{document}

\begin{frontmatter}

\cortext[cor1]{Corresponding author: \ead{mplab-me@tudelft.nl}}

\title{Unweaving the mechanics of woven fabrics through automated model discovery}

\author[inst1]{Marin~Lauber}
\author[inst1,inst2]{Uzair~Manack}
\author[inst2]{Thibault~Vervenne}
\author[inst1]{Rogier~P.~Krijnen}
\author[inst1]{\\Benjamin~Alheit}
\author[inst2]{Nele~Famaey}
\author[inst1]{Mathias~Peirlinck\corref{cor1}}

\affiliation[inst1]{organization={Department of BioMechanical Engineering, Faculty of Mechanical Engineering, Delft University of Technology (TU~Delft)},
           addressline={Delft}, 
           country={The~Netherlands}
}
\affiliation[inst2]{organization={Division of Biomechanics, Department of Mechanical Engineering, Katholieke Universiteit Leuven (KU~Leuven)},
                    addressline={Leuven},
                    country={Belgium}
}

\begin{abstract}
Woven fabrics have multiple structural, aesthetic, and medical applications that require knowledge of their mechanical properties.
While isolated fiber properties are easily determined, bulk properties of woven material, which can be coated or locally reinforced, require knowledge of yarn density, weave style, and other characteristics that are difficult to obtain experimentally.
Additionally, fiber rearrangement, twist, yarn contact, and fiber-direction coupling result in non-linear mechanical responses.
As a result, there is no universal material model for woven materials, and each fabric requires a specific model and parameter calibration.
Recently, methods to simultaneously discover material models and their parameters from experimental or synthetic datasets have been proposed.
Here, we use automated model discovery to simultaneously extract the material model and its parameters from biaxial test data of a TPU-coated woven nylon fabric.
We show that Constitutive Artificial Neural Networks discover a sparse three-term hyperelastic material law with one isotropic and two anisotropic contributions, reproducing the measured biaxial response with a mean $R^2=0.97$.
The discovered, comparatively large, isotropic contribution cannot be interpreted solely as the TPU coating response and likely also captures mechanical contributions not represented by the uncoupled yarn-direction invariants.
Across 20 initializations, predictive accuracy remains robust despite variations in the exact sparse representation, consistent with strong correlations between several candidate terms. 
LARS–LASSO independently recovers the same dominant constitutive contributions at approximately 100-fold lower computational cost.
These results are directly applicable to numerical finite-element models of fabric-based structures, including soft robotic devices.
More broadly, these results provide a foundation for further model discovery, improved constitutive modeling, and mechanistically informed optimization of woven fabric structures.
\end{abstract}


\end{frontmatter}

\section{Introduction}
\label{sec:intro}

Woven fabrics exhibit a broad range of mechanical behaviors driven by the different fibers used, yarn densities, and weave types \cite{Hu2004StructureFabrics}.
They are used for clothing, life-saving equipment, aesthetic structures, soft robotics \cite{Zhang2023SoftActuation}, and more recently, medical devices \cite{Cappello2018ExploitingActuators, Cianchetti2018BiomedicalRobotics, Arfaee2025TowardSystem, Arfaee2025AHeart}.
In these domains, accurate mechanical characterization of these fabrics is key to ensuring the durability, safety, and longevity of the resulting structures.
However, characterizing the bulk mechanical properties of a woven fabric is difficult because of the combined effects of fiber realignment, twist, yarn contact, and weave style.
These mechanical properties are particularly important for numerical models, where the experimentally observed response must be represented by a constitutive relation linking deformation and internal stress.
Once such a material model is known, these numerical models can help design more efficient fabric-based soft-robotics actuators \cite{Zhang2023SoftActuation, Armanini2023SoftOverview} and artificial hearts \cite{Lauber2026TheDevices, Guelcher2023balancing}. 

One route to constructing such a constitutive description is to resolve the fabric microstructure explicitly.
Homogenization approaches based on microscale representative volume elements can capture fiber interactions, slippage, and twisting \cite{Avery2021AFabrics, Asad2026AFabrics}, while mesoscopic models describe yarn-level mechanics through e.g. discrete sawtooth unit cells \cite{Erol2017ACell}.
Such approaches can provide detailed mechanical predictions, but require accurate microtomographic fabric imaging and isolated fiber constituent-level properties that may not be readily available during early-stage prototyping. 

A second route is to prescribe a macroscopic constitutive law and identify its response from mechanical tests.
Closed-form material models for woven fabrics have been proposed through strain energy density functions \cite{King2005AFabrics, Liu2005Two-dimensionalFabrics, Shanahan2008ElasticCloth, Peng2013ASimulation}.
These models provide explicit mathematical expressions for the strain energy based on the fabric's geometrical features and require careful calibration of the discrete constituents (fibers, yarns, weave style, etc.) of the woven material.
Alternatively, phenomenological fabric formulations can prescribe the directional constitutive response directly from experimentally measured warp, weft, and shear behavior \cite{Zhang2023SoftActuation}.
While these approaches can be applied at an early stage of design, the constitutive representation must still be selected before calibration or characterization.
For coated woven fabrics, this choice is not straightforward: the measured response combines contributions from the coating, the yarn families, and interactions between them, while the relative contribution of each is generally unknown. 
The resulting constitutive model is therefore sensitive to assumptions made by the modeler before parameter calibration, introducing modeler bias into how the mechanical response is decomposed and represented.

Automated material model discovery offers an interesting third route. 
Recently, such approaches have been proposed to extract interpretable material models of polymers, biological tissues, and knitted fabrics from multi-axial experimental test data  \cite{Linka2021ConstitutiveLearning, Peirlinck2024automated, Peirlinck2024univmatsub, Peirlinck2025a, Martonova2024cardiaccann, Vervenne2025CANNpulmonary, McCulloch2024AutomatedFabrics}.
Rather than prescribing a single constitutive form and subsequently calibrating its parameters, these approaches search over a library of admissible constitutive terms and identify both the functional form and its parameters directly from experimental data.
Constitutive Artificial Neural Networks (CANNs) embed continuum-mechanics constraints directly into their inputs, output, and architecture: invariant-based inputs encode objectivity and material symmetry, the strain-energy output ensures thermodynamic consistency, and constrained activation functions and network connectivity enforce physically admissible behavior and polyconvexity \cite{Linka2023ADiscovery, Linden2023}.
Through sparsity-promoting regularization, the resulting model can be reduced to a small number of active terms, interpreted, and subsequently implemented in finite-element procedures \cite{Peirlinck2025DemocratizingSubroutine, Alheit2026COMMET} for further device or material optimization.
Application of this approach to knitted fabrics \cite{McCulloch2024AutomatedFabrics} has highlighted complex coupling mechanics in warp-knitted fabrics. 
For coated woven fabrics, such an approach could provide an effective macroscopic constitutive model without requiring the contributions of the coating, yarn families, and their interactions to be prescribed beforehand.

In this work, we automatically discover the material model of a thermoplastic polyurethane (TPU)-coated nylon fabric recently used in soft artificial hearts \cite{Arfaee2025TowardSystem, Arfaee2025AHeart}.
To this end, we perform biaxial tensile tests on a cruciform sample of this woven fabric and use these data to discover a sparse constitutive law with a CANN.
We first identify which isotropic and anisotropic contributions are selected from the candidate constitutive library and examine how these contributions can be interpreted in relation to the physical structure of the coated fabric.

We then assess model identifiability from the available biaxial deformation states through sensitivity to initialization and correlations between candidate terms.
Finally, based on the structure revealed by the CANN, we investigate whether a least-angle regression approach can recover a similar material model more directly and efficiently.

\section{Methods}
\label{sec:method}

\subsection{Experimental protocol and biaxial testing}
\begin{figure*}
    \centering
    \includegraphics[width=1\linewidth]{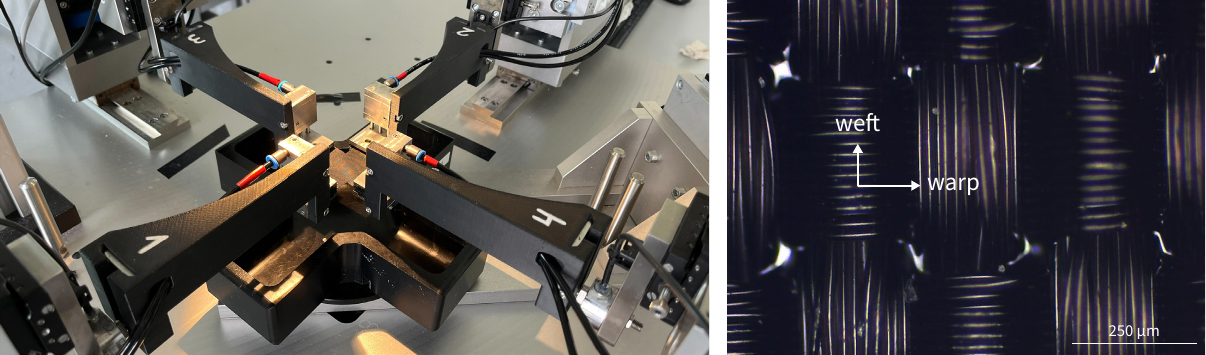}
    \caption{\textbf{Biaxial testing setup and woven architecture of the TPU-coated nylon fabric.} 
    (Left) Experimental biaxial testing setup, with the warp and weft directions aligned with the two loading axes. 
    (Right) Microscopic structure of the coated woven fabric imaged in reflection mode with a $\times10$ objective on a Leica Microsystems DM2700M. The warp direction is oriented horizontally and the weft direction vertically.
    }
    \label{fig:material}
\end{figure*}
We prepare cruciform samples of 0.2 mm thick (Riverseal 70 LW, 78 Dtex; 170 g/m²; Rivertex, Culemborg, The Netherlands) TPU-coated woven nylon fabric (see Figure~\ref{fig:material}) with an identical arm length of 20 mm and arm width of 20 mm, leaving a central square area of $20\times20$ mm to test under biaxial loading.
We use a ZwickRoell biaxial tensile tester and place the samples with the weave directions aligned with the device's testing directions.
Pneumatic clamps pressurized to 7~bar are used to hold the sample in place.
The fabric is made of two orthogonal sets of yarns, referred to as the \emph{warp} and \emph{weft} directions, which, by convention, run along and across the fabric roll, respectively (see Figure~\ref{fig:material}).
In the remainder of this paper, we also refer to these two directions as the 1- and 2-directions, respectively.
The woven nylon fabric is coated with a thin layer of TPU, making it watertight and heat-sealable.

Samples are tested in three biaxial loading conditions, similar to previous planar-biaxial protocols at multiple stretch ratios \cite{Vervenne2025pressurediameter}. 
In the equibiaxial setting, equal stretch increments are applied along the warp and weft directions. 
In the two remaining conditions, the stretch increment in the warp direction is reduced relative to that in the weft direction, resulting in ratios $(\lambda_1-1):(\lambda_2-1)$ of 1:2 and 1:3, respectively; see Table~\ref{tab:loadings}.
Each sample is tested to five maximum weft stretches, $\lambda_{2,\max}=1.02, 1.04, 1.06, 1.08,$ and $1.10$, with the corresponding $\lambda_{1,\max}$ amplitude determined by the prescribed biaxial loading condition ratio. 
Each maximum stretch amplitude is repeated for five loading–unloading cycles, resulting in a total of 25 loading cycles for each sample.
Each loading–unloading cycle starts with a reset of the loading to a preload of 0.5~N.
Once the preload is reached, we record the gauge length and reference forces and linearly increase the gripper's spacing at a rate of 0.1~mm/s.
Unloading follows an identical procedure, and the sample is unloaded until it returns to its unstretched state.

\begin{table}[h]
    \centering
    \caption{\textbf{Definition of the three biaxial loading conditions.} Minimum and maximum stretches, $\lambda_1$ and $\lambda_2$, for the three different biaxial loading conditions.}
    \begin{tabular}{lcc}
        \toprule
        experiment & max. stretch $\lambda_1$ & max. stretch $\lambda_2$ \\
        \midrule
        1:1 & 1.10 & 1.10 \\
        1:2 & 1.05 & 1.10 \\
        1:3 & 1.033 & 1.10 \\
        \bottomrule
    \end{tabular}
    \label{tab:loadings}
\end{table}

\subsection{Stress–stretch data processing}
\begin{figure*}
    \centering
    \includegraphics[width=1\linewidth]{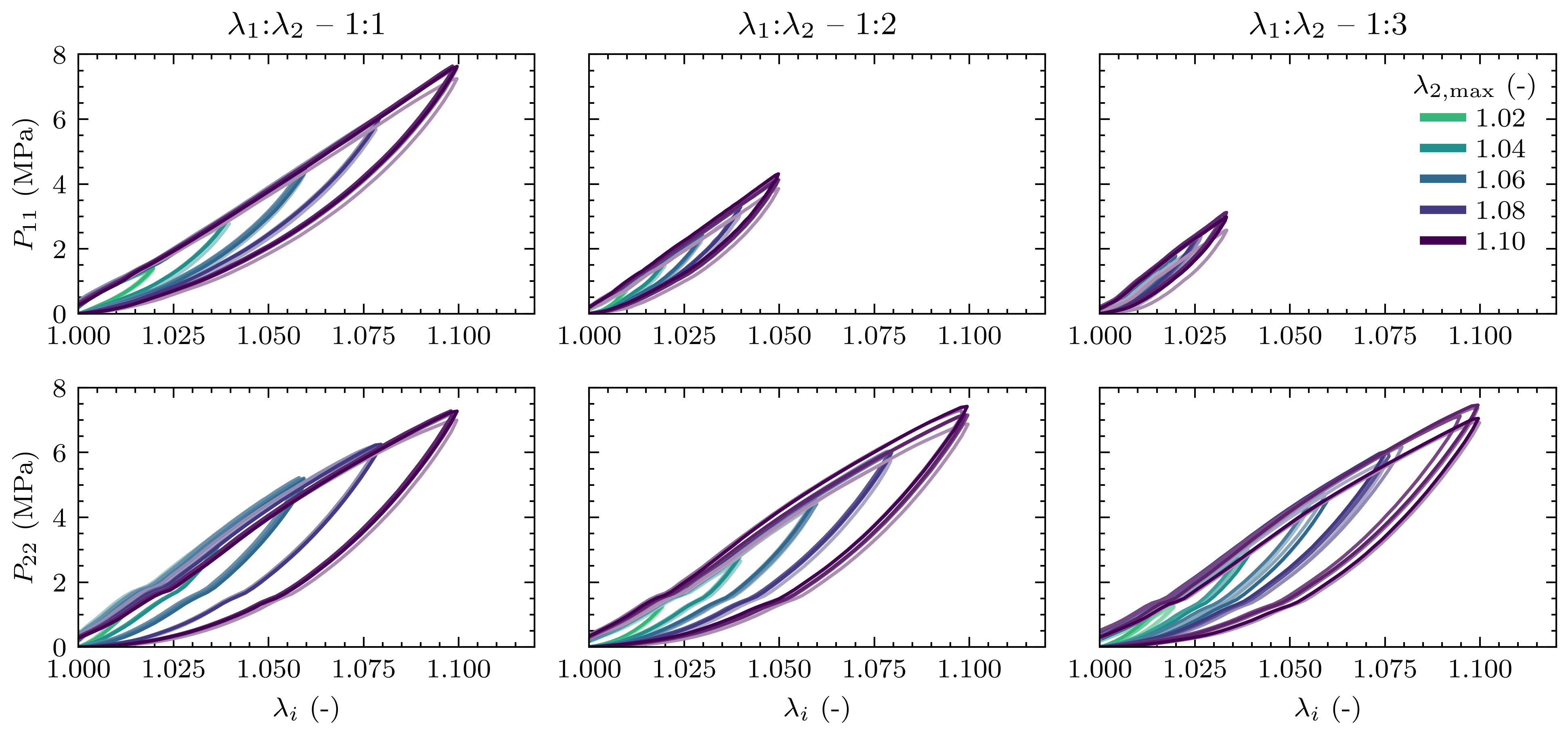}
    \caption{\textbf{Biaxial testing reveals reproducible, anisotropic, and hysteretic stress--stretch responses.}
    The non-zero first Piola--Kirchhoff stress components are shown for the three prescribed biaxial loading ratios $(\lambda_1-1):(\lambda_2-1)$ of 1:1 (left), 1:2 (center), and 1:3 (right). 
    The top and bottom rows show the warp ($P_{11}$) and weft ($P_{22}$) responses, respectively.
    Colors indicate the maximum applied weft stretch $\lambda_{2,\mathrm{max}}$, while different shades within each color show the five repeated loading--unloading cycles from light to dark.}
    \label{fig:stress-stretch}
\end{figure*}

We obtain the stress and stretch data from our force-displacement measurements by measuring the sample thickness using a Hans--Schmidt DD-50 textile thickness gauge.
To set the gauge length $L_1$ and $L_2$ and the reference forces $F_1^0$ and $F_2^0$ needed to compute the stress and stretch, we use the unloading branch of each stretch cycle.
We average the measured forces in the 1-(warp) and 2-(weft) directions to obtain $F_1^0$ and $F_2^0$, respectively, and record the values of the forces and the gripper separations when one of the forces drops below a set threshold of 0.1 N during the unloading.
We thus obtain a reference configuration for each cycle of a stretch amplitude.
This cycle-specific reference configuration reduces the influence of (accumulated) slippage effects or other artifacts that can occur during repeated loading and unloading cycles with pneumatic clamps.
Then, we compute the stretches $\lambda_1$ and $\lambda_2$ and the first Piola--Kirchhoff stresses $P_{11}$ and $P_{22}$,
\begin{equation}
\begin{split}
    \lambda_i=\frac{l_i}{L_i}, \quad P_{ii}=\frac{F_i-{F}_i^0}{t\,L_j}, \quad i=1,2; \quad j=2, 1 \,,
 \end{split}
\end{equation}
where $l_1$ and $l_2$ are the instantaneous gauge lengths, $L_1$ and $L_2$ their corresponding reference values, $F_1$ and $F_2$ the instantaneous forces obtained by averaging opposing load cells in the 1- and 2-directions, and $t$ the measured fabric thickness.
The results are five loading and five unloading curves for each of the samples and stretch amplitudes; see Figure~\ref{fig:stress-stretch}.
We resample all curves at equidistant stretch intervals and average the five repeated cycles.
For material model discovery, we use the resulting averaged unloading stress-stretch curves, providing the stress pair $\{P_{11}, P_{22}\}$ for each stretch pair $\{\lambda_1, \lambda_2\}$. 
The primary discovery analysis is performed on the three loading-ratio curves reaching $\lambda_{2,\mathrm{max}}=1.10$.

\subsection{Kinematics and invariant representation}
\label{sec:kinemtics_and_invariants}

We characterize the deformation of a material point from the reference $\vec{X}\in\Omega_0\subset\mathbb{R}^3$ to the current configuration $\vec{x}\in\Omega\subset\mathbb{R}^3$ at some time $t$ through the mapping $\vec{x} = \varphi(\vec{X},t)$.
The local deformation of the material is then obtained from the deformation gradient and right Cauchy--Green deformation tensors
\begin{equation}
    \ten{F} := \frac{\partial\varphi}{\partial\ten{X}}\,, \qquad \ten{C}:=\ten{F}^\top\ten{F} \, .
\end{equation}
To characterize the deformation of the material, we introduce two isotropic invariants, $I_1$ and $I_3$, of the right Cauchy-Green deformation tensor $\ten{C}$, and a set of anisotropic invariants $I_{4,ij}$, 
\begin{equation}
    \begin{split}
        I_1 & = \text{tr}(\ten{C}), \quad
        I_3 = \det(\ten{C}), \quad I_{4,ij}  =\ten{A}_i\cdot\ten{C}\ten{A}_j.
    \end{split}
    \label{eq:invariants}
\end{equation}
Compared to some of our prior material model discovery work, we leave $I_2$ out of our invariant set as it shows an almost perfect correlation ($r=0.9990$) with $I_1$ for all data points in our biaxial test data. 
This choice is also supported by recent work showing that the information provided by $I_1$ and $I_2$ strongly depends on the sampled deformation states \cite{Alheit2026, Dammas2025}.
In Eq.~\eqref{eq:invariants}, $\ten{A}_i$, $i=1,...,n_s$ are the $i^\text{th}$ unit structural vectors in the reference configuration. In our specific case, we have $n_s=2$, corresponding to the $\vec{n}_1$ and $\vec{n}_2$ unit warp and weft directions, respectively.

Furthermore, we assume that the fabric is incompressible, leading to $I_3=\lambda_1^2\lambda_2^2\lambda_3^2=1$.
Under a biaxial tension test, the sample stretches in two orthogonal directions, resulting in $\lambda_1\ge1$ and $\lambda_2\ge1$, and since we assume incompressibility and shear-free deformation, the deformation gradient is diagonal with $\lambda_3 =(\lambda_1\lambda_2)^{-1} \le 1$, and the first Piola--Kirchhoff stress tensor only has two non-zero components
\begin{equation}
\begin{split}
   \ten{F} &= \text{diag}\left\{\lambda_1,\lambda_2,(\lambda_1\lambda_2)^{-1}\right\} \,, \\ \ten{P} &= \text{diag}\left\{P_{11},P_{22},0\right\}.
\end{split}
\end{equation}

\subsection{Constitutive Artificial Neural Network}
We design a constitutive artificial neural network to learn the free energy density function $\psi$ for our woven material based on the invariants presented in Section~\ref{sec:kinemtics_and_invariants} and a set of network weights $\vec{\theta}=\{\theta_{(i,j,k)};\phi_{(i,j,k)}\}$, with $i=1,...,n_\mathcal{K}$, $j=1,...,n_f$, and $k=1,...,n_g$.
The network weights $\theta_{(i,j,k)}$ and $\phi_{(i,j,k)}$ are the outer and inner weights, respectively.
Here, $n_\mathcal{K}$ denotes the number of pseudo-invariants $K_i$, $n_f$ the number of polynomial functions $f_j$, and $n_g$ the number of outer functions $g_k$ included in the candidate library.
Mathematically, this is expressed as
\begin{equation}\label{eq:CANNs}
    \begin{split}
    &\psi(\ten{F};\,\vec{\theta},\ten{A}_1,...,\ten{A}_{n_s}) := \\
    &\sum_i^{n_\mathcal{K}}\sum_j^{n_f}\sum_k^{n_g}\theta_{(i,j,k)}\, g_k\left(\phi_{(i,j,k)}\,f_j\left(\mathcal{K}_i(\ten{F},\ten{A}_1,...,\ten{A}_{n_s})\right)\right)
    \end{split}
\end{equation}
where $\mathcal{K}_i$, $f_j$ and $g_k$ are pseudo-invariant and layer functions, respectively, that we describe in the following sections.

To retain an additive invariant-based representation compatible with polyconvex constitutive constructions, the network contains separate branches for each pseudo-invariant, such that the first Piola--Kirchhoff stress tensor can be additively decomposed as
\begin{equation}\label{eq:P_cann}
    \ten{P} = \frac{\partial \psi}{\partial \ten{F}} = \sum_{i}^{n_\mathcal{K}}\frac{\partial \psi}{\partial \mathcal{K}_i}\frac{\partial \mathcal{K}_i}{\partial \ten{F}} = \sum_{i}^{n_\mathcal{K}}\frac{\partial \psi}{\partial \mathcal{K}_i}\frac{\partial \mathcal{K}_i}{\partial I_i}\frac{\partial I_i}{\partial \ten{F}}\,,
\end{equation}

Motivated by these requirements and the microstructure of our material --- consisting of two in-plane fiber families of identical material and different yarn densities, coated on one side by a TPU layer -- we select the following pseudo-invariants for our neural network
\begin{equation}
\begin{split}
    \mathcal{K}_1 = I_1-3, \quad
    \mathcal{K}_{4,ij} = \begin{cases}
        \langle I_{4,ij}-1\rangle & \text{if} \quad i = j,\\
        0 & \text{otherwise},\\
    \end{cases}
\end{split}
\end{equation}
where the $\langle\circ\rangle$ Macaulay brackets restrict the yarn-associated contributions to tensile deformation. 
Cross-family invariants are not included in the present candidate library.
For each pseudo-invariant, the candidate library combines the polynomial functions
\begin{equation}
     f_j(\circ) = (\circ)^j \quad \text{for } j=\{1,3/2,2,5/2,3,4\}\,
\end{equation}
with either a linear or exponential outer function
\begin{equation}
    g_k(\circ) = \begin{cases}
        \circ & \text{if} \quad k=1,\\
        \exp(\circ) - 1 & \text{if} \quad k=2.
    \end{cases}
\end{equation}
The pseudo-invariants satisfy $\mathcal{K}_i(\ten{I})=0$ in the undeformed configuration, while the selected functions satisfy $f_j(0)=0$ and $g_k(0)=0$.
Consequently, all candidate terms vanish in the reference configuration and $\psi(\ten{I})=0$.
Furthermore, the pseudo-invariants and function families are selected following polyconvex CANN constructions \cite{Schroder2003InvariantFunctions, Alheit2026}.
For the non-negative pseudo-invariant domain considered here, the polynomial functions $f_j$ with $j\geq1$ and the linear and exponential functions $g_k$ are convex and monotonically non-decreasing.
Together with non-negative network weights, this preserves the polyconvex structure of the resulting strain-energy density.
The resulting library therefore contains both polynomial and exponential constitutive terms from which the sparse material law is discovered.

\subsection{Model training and discovery}
To discover the model and its parameters $\vec{\theta}$ that best describe our experimental data, we use the \texttt{Adam} optimizer 
to minimize a normalized least-squares error between the model-predicted stresses $\ten{P}(\ten{F}_q,\vec{\theta})$ and the corresponding experimental stresses $\hat{\ten{P}}_q$, evaluated at $q=1,...,n_\text{data}$ discrete deformation states.
We supplement this least-squares loss with an $L_p$-regularization term to promote network sparsity and improve model interpretability
\begin{equation}
\begin{split}
    L(\ten{F}; \vec{\theta}) &= \frac{1}{n_\text{data}}\sum_{q=1}^{n_\text{data}}\left\Vert\frac{\ten{P}(\ten{F}_q;\vec{\theta}) - \hat{\ten P}_q}{P_{q,\text{max}}}\right\Vert^2
    + \lambda_pL_p(\vec{\theta}).
\end{split}
\end{equation}
We define the $L_p$ norm as:
\begin{equation}
    L_p(\vec{\theta}) = \frac{1}{n_\mathcal{K}n_fn_g}\sum_i^{n_\mathcal{K}}\sum_j^{n_{f}}\sum_k^{n_{g}}\vert\theta_{(i,j,k)}\phi_{(i,j,k)}\vert^p\,.
\end{equation}
We normalize the stress residuals by the maximum stress ($P_{q,\text{max}}$) of the corresponding biaxial loading condition to ensure that each experiment is accounted for equally in the loss function.

For initialization and training, we adopt the multi-stage CANN training procedure proposed by Alheit et al. to our experimental data \cite{Alheit2026}.
We initialize all weights with positive values drawn from a uniform distribution $\vec{\theta}\sim\text{Uniform}(0,\gamma)$, where we choose $\gamma$ such that the expected area under the stress–stretch response of the randomly initialized network matches that of the measured response.
Subsequent training proceeds in three stages.
Starting with the full network, we first run 4000 epochs of the \texttt{Adam} optimizer with learning rate $\alpha_\text{Adam}=0.025$ without sparsity regularization.
Second, we continue training for 4000 epochs with $p=0.25$ and regularization weight $\lambda_p=0.01$ to promote mild sparsity.
We then prune terms whose resulting coefficient satisfies $w_i< 10^{-4}$ to construct the final set of weights.
The last phase consists of a further 4000 epochs, with an adjusted learning rate of $\alpha_\text{Adam}=0.005$, during which only the retained terms are optimized and no further sparsity is imposed.

This three-stage training strategy first lets the network fit the experimental response on a smooth optimization problem before introducing the sparsity constraint. The resulting sparse model is then fixed, and the retained coefficients are refitted without further regularization to approach their optimal values.

To assess the sensitivity of the discovered material model to network initialization, we repeat the complete three-stage training procedure for 20 independently sampled initial weight sets. 
We compare the constitutive terms retained across these runs and normalize the corresponding coefficients by the largest value obtained for each term.
To assess redundancy within the candidate library, we evaluate all candidate constitutive terms over the experimentally sampled deformation states and compute their pairwise Pearson correlation coefficients.

\subsection{Least-angle regression}\label{ssec:methods_LARS}
To investigate whether the constitutive model can be discovered more directly once the candidate library is restricted to terms linear in their coefficients, we additionally consider least-angle regression with LASSO modification (LARS-LASSO) \cite{Efron2004LeastRegression, Flaschel2026Non-smoothDiscovery}.

For this analysis, we restrict the CANN candidate library to the linear outer function $g_1(\circ)=\circ$, such that the strain-energy density becomes linear in its unknown coefficients.
Additionally, we drop the power $(\circ)^{5/2}$ as we find that this term never activates in the model discovered in the previous sections.
We then follow the LARS-LASSO material model discovery procedure proposed by Flaschel et al. \cite{Flaschel2026Non-smoothDiscovery}.
In contrast to the \texttt{Adam}-based optimization, which updates the model parameters based on gradients of the loss function, LARS constructs a sparse solution path by iteratively selecting the candidate term most strongly correlated with the current residual.
The best model is then selected as the model with the lowest BIC score across iterations (steps).
This avoids sensitivity to parameter initialization and removes the need to prescribe the regularization weight $\lambda_p$.

\section{Results}
\label{sec:results}

\subsection{Mechanical response of coated woven fabric}\label{ssec:woven-response}
Figure~\ref{fig:stress-stretch} shows the results of our biaxial tensile testing experiments.
Each row corresponds to a principal direction (warp or weft), and each column corresponds to different stretch ratios, following the test conditions defined in Table~\ref{tab:loadings}.
Within each subplot, each stretch amplitude is represented by a color from green to purple; for a given stretch amplitude, lighter shades correspond to earlier cycles and darker shades to later cycles.
All fifteen sets of experimental stress-stretch curves display a similar behavior.
Samples show a near-linear response during the loading phase; the 1-direction (warp) collapses well onto a linear elastic model, while the 2-direction response (weft) displays a slight dip in the response at $\lambda_2\sim1.02$, which is present for all stretch ratios and stretch amplitudes.
Unloading displays more non-linearity than loading, and each complete stretch cycle shows hysteresis.
One feature worth pointing out is that the dip in the force observed during loading is also present during unloading, suggesting it might originate from internal material rearrangement.
Each stretch amplitude is repeated for five cycles, and we find excellent repeatability of our experimental measurements, with all the curves collapsing onto the same hysteresis loop.
Our woven material is slightly stiffer in the warp direction than in the weft direction, which we expect since the weave structure is slightly denser in that direction; see Figure~\ref{fig:material}.
We find that peak stresses reach 7.8~MPa in the warp direction while peak stresses in the weft direction reach 7.4~MPa.

\subsection{CANN-based constitutive model discovery}\label{ssec:resultscann}
\begin{figure*}
    \centering
    \includegraphics[width=1\linewidth]{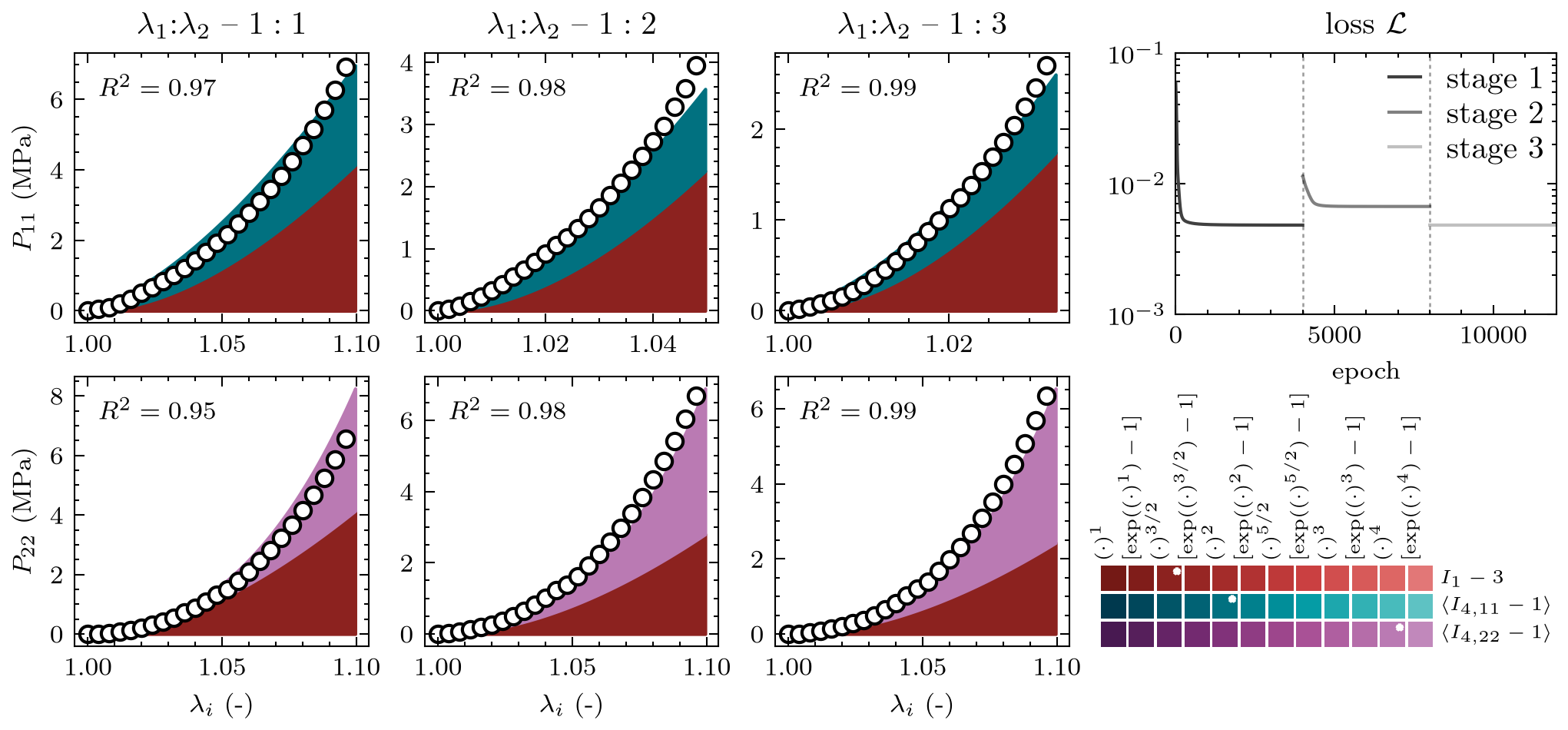}
    \caption{\textbf{CANN discovery identifies a sparse three-term constitutive model that reproduces the biaxial response.}
    The top and bottom rows show the predicted first Piola--Kirchhoff stresses in the warp ($P_{11}$) and weft ($P_{22}$) directions, respectively, for the three biaxial loading ratios.
    Experimental data are shown by circular markers, while the colored areas show the additive contributions of the isotropic and direction-specific invariants to the predicted stress.
    Different shades denote the candidate $f_j$ and $g_k$ functions associated with each invariant.
    The top-right panel shows the loss evolution over the three training stages, and the bottom-right panel summarizes the candidate library, with white stars marking the terms retained in the final model.}
    \label{fig:cann_3stage}
\end{figure*}

Our CANN discovery identifies contributions from three invariants in the final model; see Figure~\ref{fig:cann_3stage}.
Across the six stress–stretch curves, the discovered model achieves an average $R^2$ of 0.97.
Three non-zero terms remain after the pruning stage, with only minor changes to their coefficients during the final refitting stage.
The different terms are a super-linear first invariant term (Knowles-type term \cite{Knowles1977TheSolids}), a quadratic anisotropic term, and a quartic anisotropic term 
\begin{equation}\label{eq:cann_results}
\begin{split}
    \psi = & a_1(I_1-3)^{3/2} +  a_2\langle I_{4,11}-1\rangle^2 \\
    + & a_3\langle I_{4,22}-1\rangle^4.
\end{split}
\end{equation}
The stiffness-like parameters are $a_1=9.03$~MPa, $a_2=3.08$~MPa, and $a_3=51.16$~MPa.
Noteworthy, the model discovers only polynomial terms and no exponential terms, which are often associated with strain-stiffening \cite{McCulloch2024AutomatedFabrics}.

Interestingly, the isotropic contribution discovered in this model is higher than the contribution we would get following a mechanistic approach where we would fix the isotropic contribution based on the mechanical properties of the material.
Assuming a Young's modulus $E=25$~MPa and Poisson's ratio $\nu=0.33$ (Shore 90 TPU) and a coating volume fraction of $\Phi=$5\%, we obtain
\begin{equation}\label{eq:iso_fit}
    \begin{split}
    \tilde{a}_1&=\theta_{(1,1,1)}\phi_{(1,1,1)}=\Phi\mu/2\\
    &=0.05\cdot9.398/2=0.2344\text{ MPa}.
    \end{split}
\end{equation}
The much larger isotropic contribution identified by the CANN raises the question of whether part of the response attributed to this term can instead be represented by the anisotropic contributions.

\subsection{Effect of excluding isotropic contributions}
\begin{figure*}[t]
    \centering
    \includegraphics[width=1\linewidth]{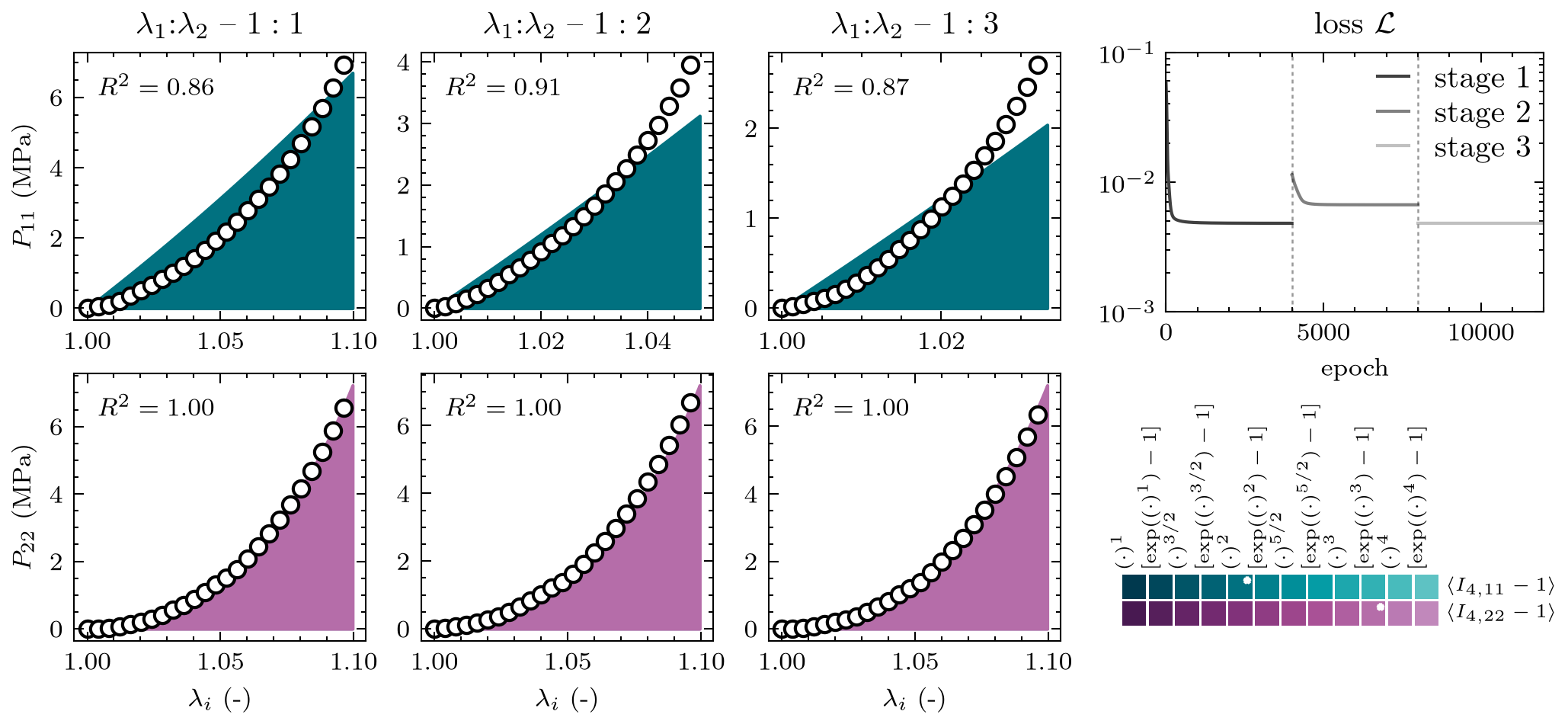}
    \caption{\textbf{Restricting the candidate library to anisotropic terms degrades the warp-direction fit.}
    The top and bottom rows show the predicted first Piola--Kirchhoff stresses in the warp ($P_{11}$) and weft ($P_{22}$) directions, respectively, after excluding all isotropic contributions from the candidate library.
    Experimental data are shown by circular markers, while the colored areas show the contributions of the two direction-specific invariants.
    Different shades denote the associated $f_j$ and $g_k$ functions.
    The top-right panel shows the loss evolution over the three training stages, and the bottom-right panel summarizes the retained constitutive terms, marked by white stars.}
    \label{fig:cann_aniso}
\end{figure*}

To investigate whether part of this isotropic contribution can be absorbed into the anisotropic terms, we repeat the model discovery using only the anisotropic invariants.
We discover a two-term model, see Figure~\ref{fig:cann_aniso}, with an average $R^2$ score of 0.94.
This $R^2$ score suggests that the model accurately captures the features of the biaxial test data.
Interestingly, the model achieves $R^2=1.00$ values in the 2-direction, while the fit in the 1-direction is poor.
The model discovered features a quadratic and a cubic exponential term
\begin{equation}\label{eq:aniso}
\begin{split}
    \psi_\text{aniso} =& a_1\langle I_{4,11}-1\rangle^2\\ +& a_2/2b_2\left[\exp(b_2\langle I_{4,22}-1\rangle^3)-1\right],
\end{split}
\end{equation}
where the two stiffness-like parameters are $a_1=7.28$~MPa, $a_2=1.15$~MPa and non-linear parameter $b_2=18.41$.
We note here that the second term in our model shows remarkable collinearity with the term $a_2\langle I_{4,22}-1\rangle^3$, and exchanging these terms results in an almost identical fit ($R^2$=0.94).
Intuitively, the fit of the model in the warp direction would improve if the sparsity constraint is relaxed; however, we find that this is not the case and that including more terms only marginally improves the $R^2$ scores in the warp direction.

\begin{figure*}[t]
    \centering
    \includegraphics[width=1\linewidth]{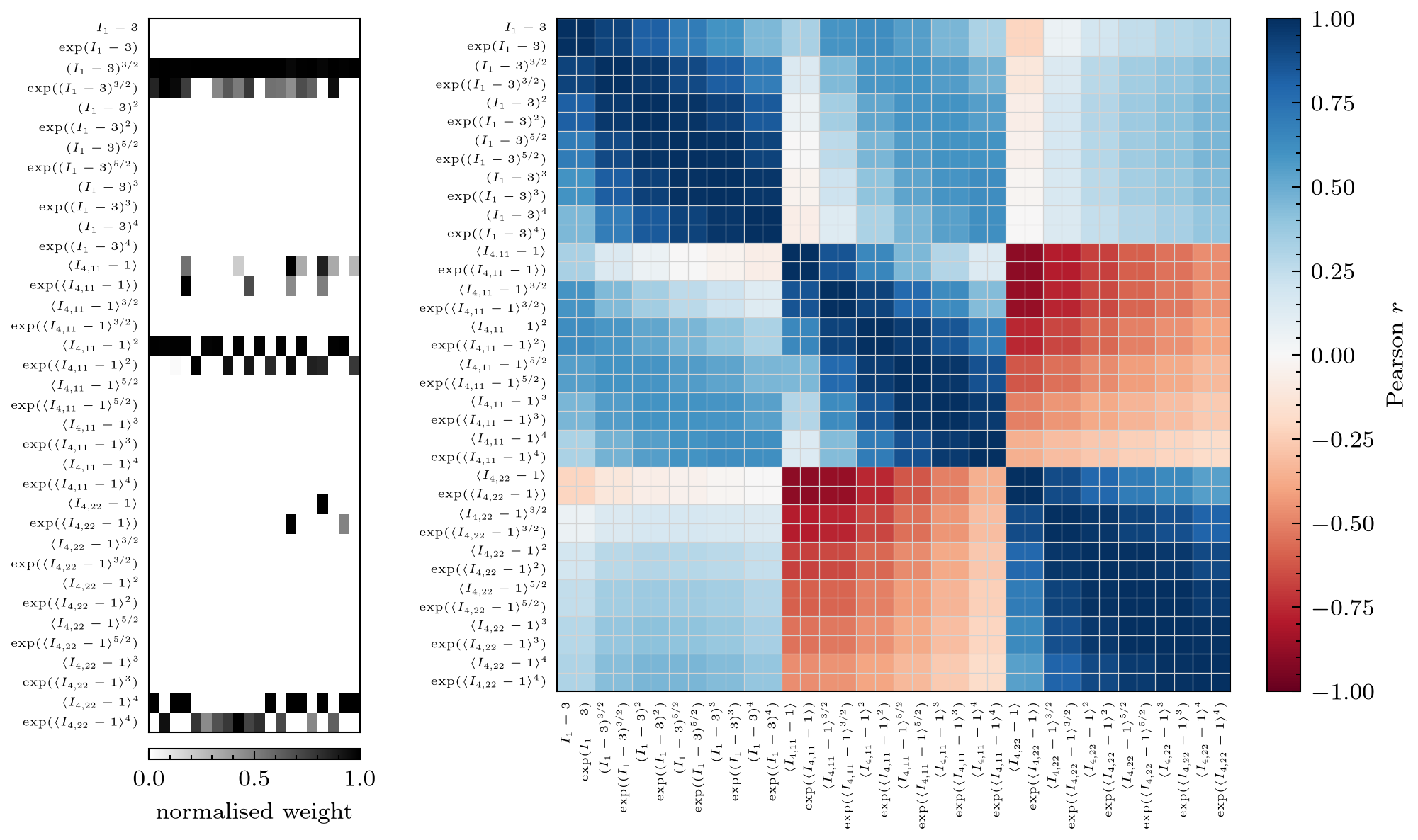}
    \caption{\textbf{Different initializations recover recurring constitutive terms, while strong candidate correlations permit alternative sparse representations.}
    (Left) Sensitivity of the discovered model to 20 independently sampled initial weight sets.
    For each candidate term, the retained coefficient is normalized by the largest value obtained for that term across all runs.
    (Right) Pairwise Pearson's $r$ correlation coefficients between the responses of all candidate constitutive terms evaluated over the experimentally sampled deformation states: each term is evaluated at the maximum seen stretch.
    Strong correlations are observed between several candidate terms, particularly between corresponding linear and exponential representations.}
    \label{fig:sensitivity}
\end{figure*}

\subsection{Model robustness and candidate-term correlations}\label{ssec:robust_corrs}

Across the 20 initializations, Figure~\ref{fig:sensitivity} -- left shows recurring constitutive terms with similar weights, although the exact sparse representation varies between runs.
The discovered models feature between 3 and 7 terms, with an average of 4 terms.
The $R^2$-score of all the models is higher than 0.97, indicating that the predictive response is robust to initialization despite variations in the selected representation.
Consistent with the strong correlation between the linear $g_1(\circ)=(\circ)$ and exponential layer function $g_2(\circ)=\exp(\circ)-1$ representations over the tested stretch range, as shown in Figure~\ref{fig:sensitivity} -- right, the discovered models find contributions from either the linear or the exponential forms of the invariants depending on the initial conditions.

The high correlation between the linear and exponential functions used to construct our constitutive neural network suggests that models of similar quality can be obtained by considering only linear functions.
This motivates the LARS–LASSO analysis considered next.

\subsection{Least angle regression -- LARS}\label{ssec:lars}

\begin{figure*}[t]
    \centering
    \includegraphics[width=1\linewidth]{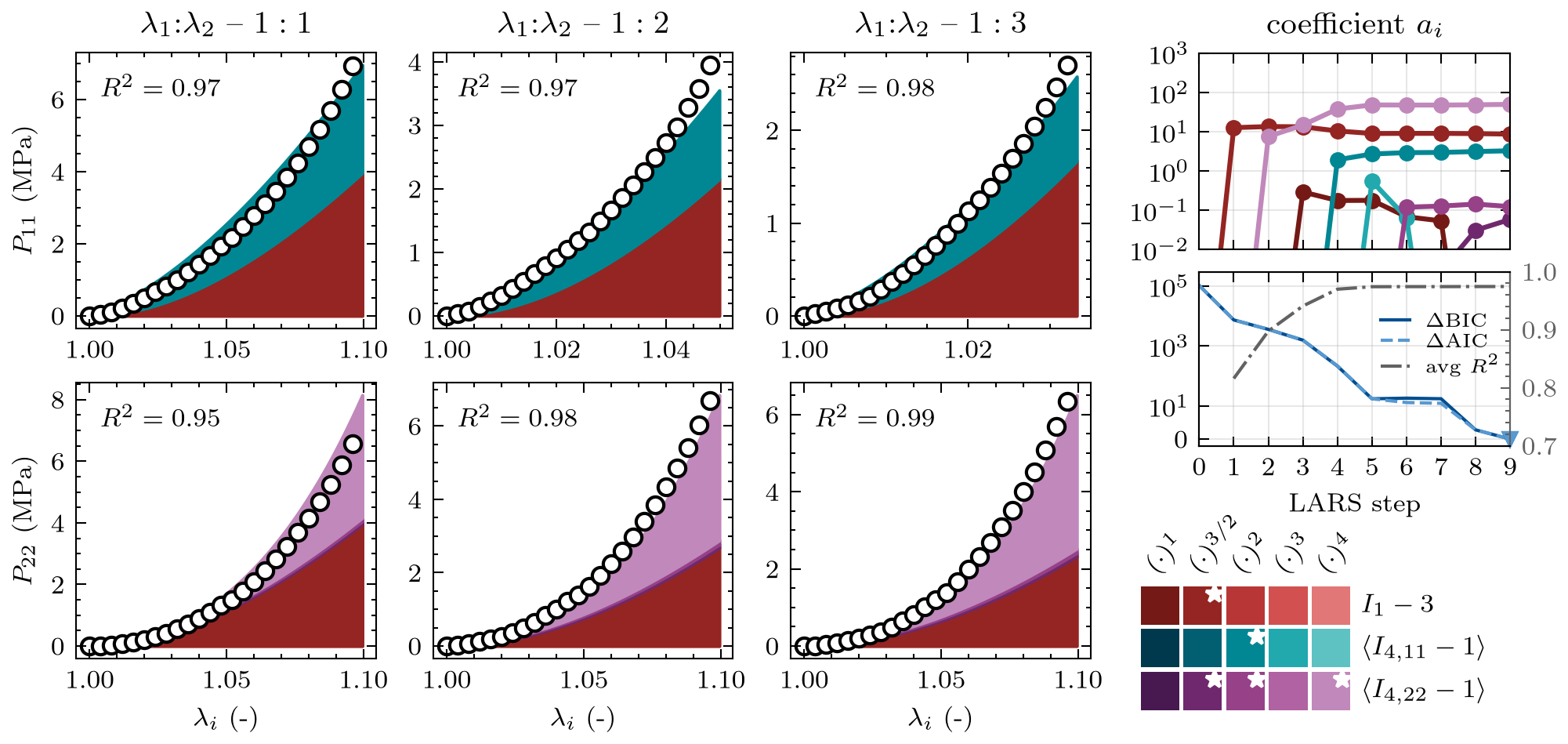}
     \caption{\textbf{LARS--LASSO recovers a closely related sparse constitutive model from the linear-in-parameter candidate library.}
    The top and bottom rows show the predicted first Piola--Kirchhoff stresses in the warp ($P_{11}$) and weft ($P_{22}$) directions, respectively, for the three biaxial loading ratios.
    Experimental data are shown by circular markers, while the colored areas show the additive contributions of the retained isotropic and direction-specific terms.
    The top-right panel shows the evolution of the constitutive coefficients along the LARS solution path, i.e. the evolution of which term and its value when it enters the solution as the algorithm progresses, together with the value of the $R^2$-score as each term enters the model (gray dash-dotted line).
    The center-right panel shows the AIC/BIC criterion values as terms enter or leave the model.
    The bottom-right panel summarizes the candidate library, with white stars marking the terms retained in the selected model.}
    \label{fig:lars}
\end{figure*}

Motivated by the results of Sections~\ref{ssec:resultscann} and \ref{ssec:robust_corrs}, we apply LARS–LASSO to the linear-in-parameter candidate library described in Sect.~\ref{ssec:methods_LARS}. 
For the present problem, LARS is approximately 100 times faster than the Adam-based CANN training.

The LARS--LASSO approach discovers a 5-term model at step 9 (top-right panel) with an average $R^2$ score of 0.974, suggesting that, again, the model captures the data well.
This model closely resembles the model discovered from the mixed linear- and nonlinear-in-parameter CANN libraries with the Adam optimizer, retaining the same super-linear isotropic $I_{1}$, quadratic $I_{4,11}$ and quartic $I_{4,22}$ contributions, together with small additional quadratic $I_{4,22}$ and $I^{3/2}_{4,22}$ invariant terms:
\begin{equation}\label{eq:LARS}
\begin{split}
    \psi_\text{LARS} =& a_1(I_1-3)^{3/2} + a_2\langle I_{4,11}-1\rangle ^2 \\
    +& a_3\langle I_{4,22}-1 \rangle^{4} + a_4\langle I_{4,22}-1 \rangle^{2}\\
    +& a_5\langle I_{4,22}-1 \rangle^{3/2}.
    \end{split}
\end{equation}
The stiffness-like parameters take the values $a_1=8.65$~MPa, $a_2=3.25$~MPa, $a_3=49.52$~MPa, $a_4=0.12$~MPa and $a_5=0.06$~MPa.
We have arranged the first three terms in Eq.~\ref{eq:LARS} in the same order as the results from Eq.~\ref{eq:cann_results} to simplify comparison since these terms are identical between the models.
The two additional terms ($a_4\langle I_{4,22}-1 \rangle^{2}$ and $a_5\langle I_{4,22}-1 \rangle^{3/2}$) contribute only a minor amount to the total strain energy; keeping only the three dominant terms reduces the $ R^2$-score to 0.973 only.
LARS discovers these dominant terms from step 5 of the training, and the inclusion of the later terms (see Figure~\ref{fig:lars} -- center right) only slightly improves the model's fit to the biaxial test data. 

\section{Discussion}
\label{sec:discussion}
\textbf{Coated woven fabric shows anisotropic non-linear responses.}
Our experimental biaxial test data show that the TPU-coated nylon fabric displays non-linear stress-stretch behavior and a slight anisotropic response, with the warp direction being stiffer than the weft.
We find excellent repeatability of the tests for the three different loading conditions as shown in Table~\ref{tab:loadings} and Figure~\ref{fig:stress-stretch}.
We discover a material model for this material that reflects both its anisotropy and the different material stiffness in the two principal directions.
However, we also observe that a very large isotropic contribution is present, much larger than the contribution we would get in a structurally-informed material model for a TPU coating (see Section~\ref{ssec:resultscann}).
We believe that this contribution absorbs response components that cannot be represented by the uncoupled warp and weft invariants in the present CANN, including potential coupling effects between the two principal weave directions.
Consistent with this interpretation, excluding all isotropic terms leaves the weft response almost unaffected but noticeably degrades the warp-direction fit; see Figure~\ref{fig:cann_aniso}. 
The isotropic contribution should therefore not be interpreted solely as a direct representation of the TPU coating.
Direct biaxial testing at $\pm45^\circ$ was not included here because cutting the present woven fabric off-axis would interrupt yarn continuity and alter the specimen architecture. 
Future studies could instead augment the data with picture-frame (trellis) tests, or with alternative biaxial specimen designs that access off-axis deformation without disrupting yarn continuity, to better identify coupling contributions.
More generally, the present experiments characterize only the in-plane (membrane) response.
Fabric-based structures \cite{Lauber2026TheDevices} can also undergo combined bending and membrane deformations, and the membrane and bending responses of woven materials may be partly decoupled
\cite{Teng1999AFabrics}. 
Characterization of this behavior would therefore require additional, dedicated experiments.

\textbf{CANNs discover complex material models for biaxial test of woven fabric.}
Our constitutive neural network discovered a three-term model for the complex non-linear anisotropic mechanical properties of the woven fabric.
As shown in Section~\ref{sec:results}, this model features one isotropic $I_1$ invariant, and two anisotropic invariants $I_{4,11}$ and $I_{4,22}$.
These types of isotropic terms are often found in hyperelastic formulations of anisotropic materials where the anisotropic contribution is embedded inside a polymeric or biological matrix; see, for example, \cite{McCulloch2024AutomatedFabrics}.
While the model captures the biaxial test data we used for the model discovery, one limitation of all the models discovered is that they are hyperelastic by design and cannot predict the amount of energy dissipated during cyclic loading.
However, from our experimental data, we are confident that this energy is small and thus hyperelastic materials offer a solid approach to model these types of fabrics.

\textbf{Exponential terms are absent from most discovered models.}
Historically, exponential anisotropic contributions have been identified as a hallmark of anisotropic soft biological tissues \cite{Holzapfel2000AModels}.
These terms have also found their way into hyperelastic formulations for knitted fabrics \cite{McCulloch2024AutomatedFabrics} and other materials \cite{Alheit2026}, where their non-linear stiffening describes well the underlying mechanics of the material.
Interestingly, our network finds exponential contributions when isotropic terms are absent from the formulation; see Figure~\ref{fig:cann_aniso}.
When exponential terms are present, they can often be exchanged with the corresponding linear term with minor changes to the overall $R^2$ score of the model; see Section~\ref{ssec:resultscann}.
Over the application-relevant stretch range probed here, which remains modest because of the high stiffness of the fabric, the linear and exponential candidate functions remain very similar, and the experimental data therefore do not strongly distinguish between these representations.
This is reflected by the high correlation of both $g_k$ functions for our stretch range; see Figure~\ref{fig:sensitivity}-right, where we consistently find almost perfect correlation scores $r>0.99$ between the linear and the exponential invariant contribution.
This also explains why different initializations can select different members of these correlated candidate families while producing almost identical stress predictions as shown in Figure~\ref{fig:sensitivity}-left: the predictive response is robust, but the exact sparse representation is not uniquely identified by the present biaxial data.

\textbf{Least angle regression offers an efficient alternative method for linear parameter models.}
Since our neural network discovered a material model with terms purely linear in the parameters, other regression techniques can be applied to our biaxial test data.
One such method that we tested here is the least-angle regression (LARS) \cite{Efron2004LeastRegression}.
Compared to CANNs, the sparsity in LARS is not imposed through an additional term in the loss function, but rather through a different algorithm.
We find that LARS discovers a material model closely related to the model we discover from the non-linear CANN basis with the classical Adam optimizer, with two additional terms with small coefficients.
In particular, the three dominant contributions identified by the CANN are recovered with similar coefficients, while the additional terms are two powered $I_{4,22}$ contributions that contribute only modestly to the total strain energy. 
This close agreement in predictive response despite small differences in the selected basis is consistent with the candidate-term correlations discussed above.
%
At the same time, the larger super-linear isotropic contribution remains in the model. This further suggests that the isotropic part of the discovered constitutive law cannot be interpreted one-to-one as the TPU coating contribution, but also captures mechanical response that is not represented by the present direction-specific invariants.

The advantage of LARS over CANN here is that there is no user-defined $\lambda_p$ term for the sparsity and that the model starts with zero coefficients and solves a convex problem; it is insensitive to initialization \cite{Flaschel2026Non-smoothDiscovery}.
However, this approach can only be applied to models linear in their parameters, and as such limits the possible models discovered.

\section{Conclusion}
\label{sec:conclusion}

In this work, we apply Constitutive Artificial Neural Networks (CANN) to discover a material model for TPU-coated woven nylon fabric directly from planar biaxial tensile test data. 
The CANN identifies a sparse three-term model consisting of a super-linear isotropic contribution and quadratic and quartic anisotropic contributions associated with the two principal yarn directions, reproducing the measured biaxial response with a mean $R^2 = 0.97$. 
The discovered, comparatively large, isotropic contribution cannot be interpreted solely as the isotropic TPU coating response and likely also captures mechanical contributions that are not represented by the uncoupled warp and weft invariants used here.
Across 20 initializations, predictive accuracy remains robust, while the exact sparse representation varies because several candidate terms are strongly correlated.
Additionally, we find that the model discovered by the CANN is linear in its parameters. Restricting the library to terms linear in their coefficients enables LARS–LASSO, which recovers the same dominant contributions at roughly 100-fold lower computational cost and avoids the need for user-defined regularization hyperparameter tuning.
Collectively, our results demonstrate automated discovery of compact constitutive laws for coated woven fabrics without prescribing the model form a priori.

\section*{Acknowledgments}
The authors gratefully acknowledge Iva van der Burgh-Vrees, Sil Gussenhoven, and Carlijn Bouten for their contributions and discussions during the early experimental development of this work, and Nienke Reitsma and Johannes T.B. Overvelde for providing the microscopic image of the coated woven fabric.

The contributions of M. Lauber were supported by the project Holland Hybrid Heart with file number NWA.1518.22.049 of the research program Onderzoek op Routes door Consortia 2022 – NWA-ORC 2022, which is financed by the Dutch Research Council (NWO), the Dutch Ministry of Education, Culture and Science (OCW), and the Hartstichting (Dutch Heart Foundation). 
The contributions of U. Manack were supported by the European Union’s Horizon Europe research and innovation programme under the Marie Skłodowska-Curie Action HORIZON-MSCA-2023-DN-01 InSilicoHealth, grant agreement No. 101169278. Views and opinions expressed are those of the authors only and do not necessarily reflect those of the European Union. Neither the European Union nor the granting authority can be held responsible for them.



\bibliographystyle{elsarticle-num} 
\bibliography{references}

\end{document}